\documentclass[a4paper]{article}

\usepackage[english]{babel}
\usepackage{url}
\usepackage{amsmath}
\usepackage{amssymb}
\usepackage{listings}
\usepackage{graphicx}
\usepackage{algorithm}
\usepackage{caption}
\usepackage{subcaption}
\usepackage[margin=1.5in]{geometry}

\title{Alignment- and reference-free phylogenomics\newline with colored de-Bruijn graphs}

\author{Roland Wittler}

\date{%
    \small Genome Informatics, Faculty of Technology, Bielefeld University, Germany\\ Center for Biotechnology, Bielefeld University, Germany\\ roland.wittler@uni-bielefeld.de\\[2ex]%
    \today
}

\newtheorem{definition}{Definition}

\newcommand{\node}[3]{\protect{\fbox{\begin{tabular}{c}#1 \\ #2\\\hline $\{#3\}$ \end{tabular}}}}

\newcommand{\uplrarrow}{\begin{picture}(0,0)\put(0,0){$\nearrow$}\end{picture}$\swarrow$}

\newcommand{\downlrarrow}{\begin{picture}(0,0)\put(0,0){$\searrow$}\end{picture}$\nwarrow$}
\begin{document}

\maketitle

\begin{abstract}
We present a new whole-genome based approach to infer large-scale phylogenies that is alignment- and reference-free. In contrast to other methods, it does not rely on pairwise comparisons to determine distances to infer edges in a tree. Instead, a colored de-Bruijn graph is constructed, and information on common subsequences is extracted to infer phylogenetic splits.
Application to different datasets confirms robustness of the approach. A comparison to other state-of-the-art whole-genome based methods indicates comparable or higher accuracy and efficiency.
\end{abstract}

\section{Introduction}



A common task in comparative genomics is the reconstruction of the evolutionary relationships of species or other taxonomic entities, their \emph{phylogeny}.
Today's wealth of available genome data enables large-scale comparative studies, where phylogenetics is faced with the following problems:
%
First, the sequencing procedure itself is becoming cheaper and faster, but finishing a genome sequence remains a laborious step. Thus, more and more genomes are published in an unfinished state, i.e., only assemblies (composed of contigs), or raw sequencing data (composed of read sequences) are available.
Hence, traditional approaches for phylogenetic inference can often not be applied, because they are based on the identification and comparison of marker sequences, which relies on computing multiple alignments, an NP-hard task.
Second, the low sequencing cost allow new large-scale studies of certain niches and/or aloof from model organisms, where reference sequences would be too distant or not available at all.

Whole-genome approaches solve these problems as they are usually alignment- and reference-free, see e.g.~\cite{aaf,andi,fswm,co-phylog,swphylo,cvtree3}
. However, the sheer number of genomes to be analysed is still posing limits in large-scale scenarios as
almost all whole-genome approaches are based on a pairwise comparison of some characteristics of the genomes (e.g.\ occurrences or frequencies of $k$-mers or other patterns) to define distances which are then used to reconstruct a tree (e.g.\ by using neighbor joining~\cite{nj}). This means, for $n$ genomes, $O(n^2)$ comparisons are performed in order to infer $O(n)$ edges.
To the best of our knowledge, only MultiSpaM~\cite{multispam} follows a different approach by sampling small, gap-free alignments involving \emph{four} genomes each, which are used to infer a super tree on quartets. According to our experiments, this method is not suitable for large-scale settings (see Results), though.

Apart from computational issues, the actual objective of phylogenetic inference in terms of how to represent a phylogeny is not obvious in the first place.
Taking only intra-genomic mutations into account, i.e., assuming a genome mutating independently of others, genomes would have unique lines of ancestors and their phylogeny would thus be a tree.
Several reasons however conflict this simple tree model. Inter-genomic exchange of genomic segments such as crossover in diploid or polyploid organisms, lateral gene transfer in bacteria, or introgression in insects contradict the assumption of unique ancestry. Furthermore, incomplete, ambiguous, or even misleading information can hamper resolving a reliable phylogenetic tree.

\medskip

Here, we propose a new methodology that is whole-genome based, alignment- and reference-free, and does not rely on a pairwise comparison of the genomes or their characteristics.
An implementation called SANS (''Symmetric Alignment-free phylogeNomic Splits``) is available at \url{https://gitlab.ub.uni-bielefeld.de/roland.wittler/sans}.
The $k$-mers of all genomic sequences (assemblies or reads) are stored in a colored de-Bruijn graph, which is then traversed to extract phylogenetic signals.
The reconstructed phylogenies are not restricted to trees. Instead, the generalized model of \emph{phylogenetic splits}
~\cite{bandeltdress} is used to infer phylogenetic networks that can indicate a tree structure and also point to ambiguity in the reconstruction.

\medskip
In the following Section~\ref{sec:background}, we will first introduce two building blocks of our approach, \emph{splits} and \emph{colored de-Bruijn graphs}. Then, we will describe and motivate our method in Section~\ref{sec:method}. After an evaluation on several real data sets in Section~\ref{sec:results}, we will give a brief summary and an outlook in Section~\ref{sec:conclusion}.

\section{Background}
\label{sec:background}

Before presenting our method in Section~\ref{sec:method}, we will introduce two basic concepts it builds upon. Firstly, as motivated above, our phylogenies will be represented by sets of \emph{splits}, a generalization of trees. Secondly, to extract phylogenetic signals from the given genomes at the first place, they are stored in a \emph{colored de-Bruijn graph}.

\subsection{Phylogenetic splits}
\label{sec:splits}

In the following, we briefly recapitulate some notions and statements from the split decomposition theory introduced by Bandelt and Dress~\cite{bandeltdress}, and put them into context.


\begin{definition}[Unordered split]\label{def:split}
Given a set $O$, if for two subsets $A,B\subseteq O$, both $A\cap B=\emptyset$ and $A\cup B=O$, then the unordered pair $\{A,B\}$ is a \emph{bipartition} or \emph{(unordered) split} of $O$.
If either $A$ or $B$ is empty, a split is called \emph{trivial}.
\end{definition}

We extend the above commonly used terminology of (unordered) splits to \textit{ordered splits}---a central concept in our approach.

\begin{definition}[Ordered split]\label{def:orderedSplit}
If $\{A,B\}$ is an unordered split of $O$, the ordered pairs $(A,B)$ and $(B,A)$ are \emph{ordered splits}. $(B,A)$ is called the \emph{inverse (split)} of $(A,B)$ and \textit{vice versa}.
\end{definition}

Note that one unordered split $\{A,B\}=\{B,A\}$ corresponds to two ordered splits $(A,B) \neq (B,A)$. Our method will first infer ordered splits and their inverse, which will then be combined to form unordered splits.
If clear from the context, we may denote an ordered split $(A,B)$ by simply $A$.

A set of splits $\mathcal{S}$ may be supplemented with weights $w:\mathcal{S}\longrightarrow\mathbb{R}$, e.g., in~\cite{bandeltdress}, splits are weighted by a so-called \emph{isolation index}.
Strong relations between metrics and sets of weighted unordered splits have been shown. In particular,  one can derive a unique set of weighted splits $\mathcal{S}_d$ from any metric distance $d$ such that $d(a,b) = \sum_{\{A,B\} \in \mathcal{S}}{w(\{A,B\}) \, \delta_{A}(a,b)}$ where $\delta_{A}(a,b):=1$ if either $a$ or $b$ in $A$, but not both, and $\delta_{A}(a,b):=0$ otherwise, i.e., the weights of all splits which separate $a$ from $b$ are accumulated. A set of splits $\mathcal{S}$ is of the form $\mathcal{S}=\mathcal{S}_d$ for some metric $d$ if and only if it is \emph{weakly compatible} in the following sense.

\begin{definition}[Weak compatibility~\cite{bandeltdress}]\label{def:weaklycompatible}
 A set of unordered splits $\mathcal{S}$ on $O$ is \emph{weakly compatible} if for any three splits $\{A_1,B_1\}$, $\{A_2,B_2\}$, $\{A_3,B_3\}$ $\in \mathcal{S}$, there are no elements $a, a_1, a_2, a_3 \in O$ with $\{a,a_1, a_2, a_3\}\cap A_i = \{a,a_i\}$ for $i=1,2,3$.
\end{definition}

As a peculiarity of our approach is being \emph{not} distance-based, we mention the above relation of weakly compatible splits and distances only for the sake of completeness. We will make use the above property to filter a general set of splits such that it can be displayed as a---in most cases planar---network.

For a tree metric (also called \emph{additive metric})~$d$, a set of splits $\mathcal{S}$ is of the form $\mathcal{S}=\mathcal{S}_d$, if and only if it is \emph{compatible} in the following sense.

\begin{definition}[Compatibility~\cite{bandeltdress}]\label{def:compatible}
 A set of unordered splits $\mathcal{S}$ on $O$ is \emph{compatible} if for any two splits $\{A,B\}$ and $\{A',B'\}$, one of the four intersections $A\cap A'$, $A\cap B'$, $B\cap A'$, and $B\cap B'$ is empty.
\end{definition}

We will make use of the implied one to one correspondence of edges in a tree and compatible splits: an edge of length~$w$ whose removal separates a tree into two trees with leaf sets~$A$ and~$B$, respectively, corresponds to a split $\{A,B\}$ of weight~$w$.

\subsection{Colored de-Bruijn graphs}
\label{sec:C-DBG}

A \emph{string}~$s$ is a sequence of characters over a finite, non-empty set, called \emph{alphabet}. Its \emph{length} is denoted by~$|s|$, the character at position~$i$ by~$s[i]$, and the substring from position~$i$ through~$j$ by $s[i..j]$. A substring of length $k$ is called \emph{$k$-mer}.

We consider a \emph{genome} as a set of strings over the DNA-alphabet $\{A,C,G,T\}$. The \emph{reverse complement} of a string $s$ is $\overline{s}:=\overline{s[|s|]}\cdots\overline{s[1]}$, where $\overline{A}:=T, \overline{C}:=G, \overline{G}:=C, \overline{T}:=A$.

An abstract data structure that is often used to efficiently store and process a collection of genomes is the \emph{colored de-Bruijn graph (C-DGB)}~\cite{iqbal2012novo}. It is a node-labeled graph $(V,E,col)$ where each vertex $v \in V$ represents a $k$-mer associated with a set of colors $col(v)$ representing the set of genomes the $k$-mer occurs in. A directed edge from $v$ to $v'$ exists if and only if for the corresponding $k$-mers~$x$ and~$x'$, respectively, $x[2..k] = x'[1..k-1]$.
We call a path $p=v_1,\ldots,v_l$ of length $|p|=l$ in a C-DBG \emph{non-branching} if all contained vertices have an in- and outdegree of one with the possible exception of $v_1$ having an arbitrary indegree and $v_l$ having an arbitrary outdegree, and it has the same set of colors assigned to all its vertices. A maximal non-branching path is a \emph{unitig}. In a \emph{compacted C-DBG}, all unitigs are merged into single vertices.

In practice, since a genomic sequence can be read in both directions, and the actual direction of a given sequence is usually unknown, a string and its reverse complement are assumed equivalent. Thus, in many C-DBG implementations, both a $k$-mer and its reverse complement are represented by the same vertex. In the following, we will assume this being internally handled by the data structure.

\section{Method}
\label{sec:method}

The basic idea of our new approach is that a sequence which is contained as substring in a subset~$A$ of all genomes~$G$ but not contained in any of the other genomes is interpreted as a signal that $A$ should be separated from $G\backslash A$ in the phylogeny. The more of those sequences exist and the longer they are, the stronger is the signal for separation.

To efficiently extract common sequences, we first construct a C-DBG of all given genomes. Then, we collect all separation signals as ordered splits, where any unitig~$u$ contributes $|u|$ to the weight of an ordered split $col(u)$.
Since both an ordered split $(A,B)$ and its inverse $(B,A)$ indicate that $A$ and $B$ should be separated in the phylogeny, we combine them to one unordered split $\{A,B\}$ with an overall weight that is a combination of the individual weights. The individual steps will be explained in more detail next.

\subsubsection*{C-DBG} Among several available implementations of C-DGBs (e.g.~\cite{almodaresi2017rainbowfish,holley2016bloom,iqbal2012novo,muggli2017succinct}), we decided to use Bifrost (Paul Melsted and Guillaume Holley, \url{https://github.com/pmelsted/bifrost}) for the following reasons: it is easy to install and use; it is efficiently implemented; it can process full genome sequences, assemblies, read data or even combinations of these; for read data as input, it offers some basic assembly-like filtering of $k$-mers; and it realizes a compacted C-DBG and provides a C++ API such that a traversal of the unitigs could be easily and efficiently implemented---only unitigs with heterogeneous color sets had to be split, because colors are not considered during compaction.

\subsubsection*{Accumulating split weights} Because many splits are overlapping, we use a trie data structure to store a split (as key) as path from the root to a terminal vertex, along with its weight (as value) assigned to the terminal vertex. 
We represent the set of genomes~$G$ as a list with some fixed order, and any subset of $G$ as sublist of $G$, i.e., with the same relative order.
For a split $(A,B)$ and its inverse $(B,A)$, we take as key the shorter of $A$ and $B$, breaking ties by selecting that split containing $G[0]$, and as value the pair of weights $(w,w')$, where $w$ is the accumulated weight of the key, and $w'$ the accumulated weight of its inverse.
When the trie is accessed for a key the first time, the value is initialized with $(0,0)$.

\medskip
The overall method SANS is shown in Algorithm~\ref{alg}, the very last step of which will be motivated in the following.

\begin{algorithm}[t]
\caption{\label{alg}SANS: Symmetric, Alignmet-free phylogeNomic Splits}
\begin{verbatim}
INPUT: List of genomes G
OUTPUT: Weighted splits over G
T := empty trie // initialize T[S] := (0,0) on first access by S
C-DBG := colored de-Bruijn graph of G
foreach unitig U in C-DBG:
    S := color list of U (sublist of G)
    // add ordered split S or its inverse G\S to trie
    if |S| < |G|/2 or ( |S| == |G|/2 and S[0] == G[0] ) then:
        increase first element of T[S] by length of U
    else:
        increase second element of T[G\S] by length of U
foreach entry S in T with values (w,w'):
    output unordered split {S,G\S} of weight sqrt(w*w')
\end{verbatim}
\end{algorithm}

\subsubsection*{Combining splits and their inverses} To combine an ordered split $(A,B)$ of weight~$w_A$ and its inverse $(B,A)$ of weight~$w_B$, a naive argument would be: both indicate the same separation, so they should be taken into account equivalently, and thus take the sum $w_A + w_B$ or arithmetic mean $(w_A + w_B)/2$. However, in our evaluation, this weighting scheme often assigned higher weight to wrong splits than to correct splits (compared to reliable reference trees; exemplified in Section~\ref{sec:drosophila}). Instead, we revert the above argument: 
consider a mutation on a (true) phylogenetic branch separating the set of genomes into subgroups~$A$ and~$B$. The corresponding two variants of the affected segment will induce two unitigs with color sets~$A$ and~$B$, respectively. Under the infinite sites assumption, these unitigs would not be affected by other events. So, each mutation on a branch in the phylogeny contributes to \emph{both} splits $(A,B)$ and $(B,A)$.
We hence take the geometric mean $\sqrt{w_A \cdot w_B}$ such that in case of asymmetric splits, the lower weight diminishes the total weight, and only symmetric splits receive a high overall weight.

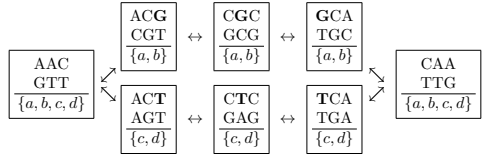
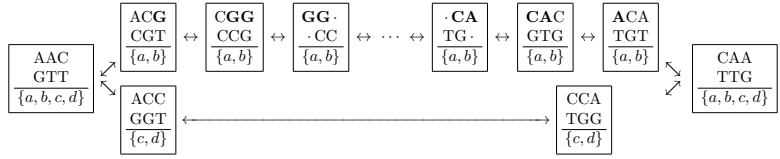
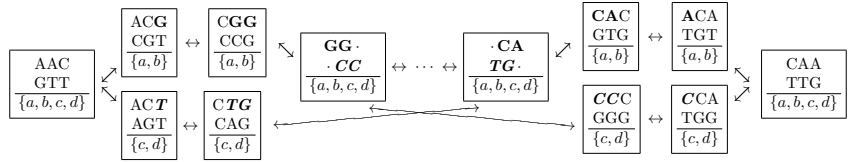
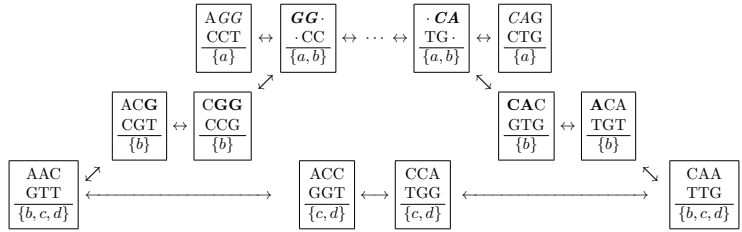
\begin{figure}
\setlength{\tabcolsep}{2pt}
\begin{subfigure}{\textwidth}
\centering
\scalebox{0.6}{
\begin{tabular}{ccccc}
 \node{AAC}{GTT}{a,b,c,d} &
  \parbox{2ex}{\uplrarrow \phantom{XXX} \downlrarrow} &
  \parbox{35ex}{\node{AC\textbf{G}}{CGT}{a,b}  $\leftrightarrow$ \node{C\textbf{G}C}{GCG}{a,b} $\leftrightarrow$ \node{\textbf{G}CA}{TGC}{a,b} \phantom{XXXXXXXXXXXXXXXXXX}  
  \node{AC\textbf{T}}{AGT}{c,d}  $\leftrightarrow$ \node{C\textbf{T}C}{GAG}{c,d}  $\leftrightarrow$  \node{\textbf{T}CA}{TGA}{c,d} \phantom{XXXXXXXXXXXXXXXXXX}} &
  \parbox{3ex}{\downlrarrow \phantom{XXX} \uplrarrow} &
  \node{CAA}{TTG}{a,b,c,d}
\end{tabular}
 }
\caption{Single nucleotide variation in genomes $a=b=\;$AAC\textbf{G}CAA and $c=d=\;$AAC\textbf{T}CAA. The induced ordered split $\{a,b\}$ and its inverse $\{c,d\}$ of weight $k=3$ each yield a corresponding unordered split $\left\{\{a,b\},\{c,d\}\right\}$ of weight $\sqrt{k\,k}=k=3$.}
\end{subfigure}

\bigskip\bigskip
\begin{subfigure}{\textwidth}
\centering
\scalebox{0.6}{
\begin{tabular}{ccccc}
  \node{AAC}{GTT}{a,b,c,d} &
  \parbox{2ex}{\uplrarrow \phantom{XXX} \downlrarrow} &
 \parbox{78ex}{\node{AC\textbf{G}}{CGT}{a,b}  $\leftrightarrow$ \node{C\textbf{GG}}{CCG}{a,b} $\leftrightarrow$ \node{\textbf{GG$\,\cdot\,$}}{$\,\cdot\,$CC}{a,b} $\leftrightarrow$ $\,\cdots$ $\leftrightarrow$ \node{\textbf{$\,\cdot\,$CA}}{TG$\,\cdot\,$}{a,b} $\leftrightarrow$ \node{\textbf{CA}C}{GTG}{a,b} $\leftrightarrow$ \node{\textbf{A}CA}{TGT}{a,b}
 \phantom{XXXXXXXXXXXXXXXXXXXXXXXXXXXXXXXXXxxXXXXXXX}
  \node{ACC}{GGT}{c,d} $\overleftrightarrow{\hspace*{8.1cm}}$ \node{CCA}{TGG}{c,d}} &
  \parbox{3ex}{\downlrarrow \phantom{XXX} \uplrarrow} &
  \node{CAA}{TTG}{a,b,c,d}
\end{tabular}
 }
 \caption{Insertion/deletion of length $l=4$ (or longer, indicated by dots) in genomes $a=b=\;$AAC\textbf{GG$\,\cdots$CA}CAA and $c=d=\;$AACCAA. The induced ordered split $\{a,b\}$ of weight $l+k-1 = l+2$ and its inverse $\{c,d\}$ of constant weight $k-1 = 2$ yield a corresponding unordered split $\left\{\{a,b\},\{c,d\}\right\}$ of weight $\sqrt{(l+k-1)\,(k-1)}=\sqrt{2(l+2)}$.}
\end{subfigure}
\centering

\bigskip\bigskip
\begin{subfigure}{\textwidth}
\centering
 \scalebox{0.6}{
\begin{tabular}{ccccccccc}
 \node{AAC}{GTT}{a,b,c,d} &
  \parbox{2ex}{\uplrarrow \phantom{XXX} \downlrarrow} &
  \parbox{21ex}{\node{AC\textbf{G}}{CGT}{a,b}  $\leftrightarrow$ \node{C\textbf{GG}}{CCG}{a,b}
    \phantom{XXXXXXXXXXXXXXXXXXX}
    \node{AC\textbf{\textit{T}}}{AGT}{c,d}  $\leftrightarrow$ \node{C\textbf{\textit{TG}}}{CAG}{c,d}}
  &
  \begin{picture}(0,0)
   \put(0,-26){\rotatebox{5}{$\overleftrightarrow{\hspace*{130pt}}$}}
  \end{picture}
  \raisebox{5ex}{\downlrarrow} 
  &
  \raisebox{2.5ex}{\node{\textbf{GG$\,\cdot\,$}}{\textbf{\textit{$\,\cdot\,$CC}}}{a,b,c,d} $\leftrightarrow$ $\,\cdots$ $\leftrightarrow$ \node{\textbf{$\,\cdot\,$CA}}{\textbf{\textit{TG$\,\cdot\,$}}}{a,b,c,d}}&
  \raisebox{5ex}{\uplrarrow}
  \begin{picture}(0,0)
   \put(-130,-15){\rotatebox{-5}{$\overleftrightarrow{\hspace*{130pt}}$}}
  \end{picture}
  &
  \parbox{21ex}{\node{\textbf{CA}C}{GTG}{a,b} $\leftrightarrow$ \node{\textbf{A}CA}{TGT}{a,b} 
  \phantom{XXXXXXXXXXXXXXXXXXX}
     \node{\textbf{\textit{CC}}C}{GGG}{c,d} $\leftrightarrow$ \node{\textbf{\textit{C}}CA}{TGG}{c,d}
  \phantom{XXXXXXXXXXXXXXXXXXXXXXXXXXXxXXXXX}} &
  \parbox{3ex}{ \downlrarrow \phantom{XXX} \uplrarrow} &
  \node{CAA}{TTG}{a,b,c,d}
\end{tabular}
 }
  \caption{Inversion of  length $l=4$ (or longer, indicated by dots) between genomes $a=b=\;$AAC\textbf{GG$\,\cdots$CA}CAA and $c=d=\;$AAC\textbf{\textit{TG$\,\cdots$CC}}CAA. The induced ordered split $\{a,b\}$ and its inverse $\{c,d\}$ of constant weight $2(k-1) = 4$ each yield a corresponding unordered split $\left\{\{a,b\},\{c,d\}\right\}$ of constant weight $\sqrt{2(k-1)\,2(k-1)}=2(k-1)=4$.}
\end{subfigure}
 
\bigskip\bigskip
\begin{subfigure}{\textwidth}
\centering
\scalebox{0.6}{
\begin{tabular}{cp{3ex}rccclp{3ex}c}
 && \node{A\textit{GG}}{CCT}{a} & $\leftrightarrow$ & \node{\textbf{\textit{GG$\,\cdot\,$}}}{$\,\cdot\,$CC}{a,b} $\leftrightarrow$ $\,\cdots$ $\leftrightarrow$ \node{\textbf{\textit{$\,\cdot\,$CA}}}{TG$\,\cdot\,$}{a,b} & $\leftrightarrow$ & \node{\textit{CA}G}{CTG}{a} &&\\
 &&&\uplrarrow &&\downlrarrow &&& \\
 &&\node{AC\textbf{G}}{CGT}{b}  $\leftrightarrow$ \node{C\textbf{GG}}{CCG}{b} &&&&\node{\textbf{CA}C}{GTG}{b} $\leftrightarrow$ \node{\textbf{A}CA}{TGT}{b} &\\
 \node{AAC}{GTT}{b,c,d} & \begin{picture}(0,0)\put(0,0){$\overleftrightarrow{\hspace*{117pt}}$}\end{picture}\raisebox{3ex}{\uplrarrow}  & 
&&\node{ACC}{GGT}{c,d} $\longleftrightarrow$ \node{CCA}{TGG}{c,d}&\begin{picture}(0,0)\put(-14,0){$\overleftrightarrow{\hspace*{117pt}}$}\end{picture}&
 &\raisebox{3ex}{\downlrarrow}  & \node{CAA}{TTG}{b,c,d}\\
\end{tabular}
}
 \caption{Lateral gene transfer of length $l=4$ (or longer, indicated by dots) from genome $a=\;$A\textit{GG$\,\cdots$CA}G to $b=\;$AAC\textbf{GG$\,\cdots$CA}CAA but not to $c=d=\;$AACCAA. Apart from mutation-independent splits for the boundaries, and the trivial split $\{b\}$ (without its inverse), the split $\{a,b\}$ of weight $l-k+1=l-2$ and its inverse $\{c,d\}$ of constant length $k-1=2$ are induced, yielding a corresponding unordered split $\left\{\{a,b\},\{c,d\}\right\}$ of weight $\sqrt{(l-k+1)\,(k-1)}=\sqrt{2(l-2)}$.}
\end{subfigure}
\caption{\label{fig:examples} Toy examples for different mutations within four genomes $a,b,c$ and $d$ to illustrate their effect on a C-DBG with $k=3$. Each vertex of the C-DBG is labelled with both its $k$-mer and the reverse complement (in arbitrary order), as well as its color set. Due to the small value of $k$, the C-DBG contains edges corresponding to pairs of overlapping $k$-mers that are not contained in the given strings. For the purpose of clarity, these are not drawn. Mutations are highlighted in bold and/or italics.}
 \end{figure}

Considering different scenarios that would affect the observation of common substrings in the C-DBG, some of which are illustrated in Figure~\ref{fig:examples}, we observe beneficial behavior of the weighting scheme in almost all cases:
A \textbf{single nucleotide variation} would cause a bubble in the C-DBG composed of two unitigs of similar length~$k$ each---a symmetric scenario in accordance with the above weighting scheme.
Both an \textbf{insertion or deletion} of length~$l$ would cause an asymmetric bubble and thus asymmetric weights $k-1$ and $l+k-1$. Here, the geometric mean has the positive effect to weaken the impact of the length of the event on the overall split weight.
For both a \textbf{transposition or inversion} of arbitrary length, the color set of the segment itself remains the same, and only those $k$-mers spanning the breakpoint regions would be affected, inducing symmetric bubbles in accordance with the  weighting scheme.
\textbf{Lateral gene transfer} is challenging phylogenetic reconstruction, because a subsequence of length~$l$ that is contained in both the group~$A$ containing the donor genome as well as the target genome~$b$ from the other genomes $B:= G \backslash A$ can easily be misinterpreted as a signal to separate $A\cup\{b\}$ from the remainder~$B\backslash\{b\}$ instead of separating $A$ from $B$, where the strength of this erroneous signal grows with $l$. Our approach will be affected only little: On the one hand, the unitig corresponding to the copied subsequence has color set $A\cup\{b\}$ and thus contributes to an ordered split $(A\cup\{b\},B\backslash\{b\})$ of weight~$l-k+1$. On the other hand, because the transfer does not remove any subsequence in the donor sequence, only those $k$ $k$-mers spanning the breakpoint region will be affected, inducing a unitig with color set $B\backslash\{b\}$ whose length is independent of $l$.
\textbf{Missing or additional data} may arise from genomic segments that are difficult to sequence or assemble and might thus be missing in some assemblies, due to the usage of different sequencing protocols, assembly tools, or filter criteria, or simply because some input files contain plasmid or mitochondrial sequences and others do not. This does not affect our approach, because additional sequence induces unitigs and thus an ordered split, but the absence of sequence does not induce any split, not even due to breakpoint regions, because in such cases usually whole reads, contigs or chromosomes are involved. Thus, the weight of the additional ordered split would be multiplied by zero for the absent split, resulting in a total weight of zero.
\textbf{Copy number changes} can only be detected if the change is from one to two or \textit{vice versa}, adding or removing $k$-mers spanning the juncture of the two copies. Beyond that, because the $k$-mer counts are not captured, our approach is not sensible for copy number changes.

In practice, the structure of a C-DBG is much more complex than the simplified picture we draw above. Nevertheless, using the geometric yields high accuracy of the approach compared to other methods.

\subsubsection*{Postprocessing}

Even though the geometric mean filters out many asymmetric splits, the total number of positively weighted splits can be many-fold higher than $2n-3$, the number of edges in a fully resolved tree for $n$ genomes. Unfortunately, the observed distribution of split weights did not indicate any obvious threshold to separate high-weighted splits from low-weighted noise. Nevertheless, a rough cutoff can safely be applied by keeping only the $t$ highest weighting splits, e.g., in our evaluation $t=10\,n$ has been used for all datasets.
Additionally, we evaluated two filtering approaches: 
\emph{greedy weakly}, i.e., greedily approximating a maximum weight subset that is weakly compatible and can thus be displayed as a network, and
\emph{greedy tree}, i.e., greedily approximating a maximum weight subset that is compatible and thus corresponds to a tree. To this end, we used the corresponding options of the software tool SplitsTree~\cite{splitstree_an,splitstree}. As we will demonstrate in the Results section, in particular the tree filter proved to be very effective in practice.

\subsubsection*{Run time complexity}
Consider $n$ genomes of length $O(m)$ each. 
In Bifrost, the compacted C-DBG is built by indexing a $k$-mer by its \emph{minimizer}, i.e., a substring with the smallest hash value
among all substrings of length $g$ in a $k$-mer. According to the developers of Bifrost (personal communication), inserting a $k$-mer and its color takes $O(4^{(k-g)} log(n))$ time in the worst case.
In practice, however, each of the $O(m\,n)$ $k$-mers can be inserted in $O(\log(n))$ time, and hence, building the complete C-DGB takes $O(m\,n\log(n))$ time.
While iterating over all positions in the graph, we verify whether a unitig has to be split due to a change in the color set. Because each of the $n$ genomes adds $O(m)$ color assignments to the graph, we have to do $O(m\,n)$ color comparisons in total, which does not increase the overall complexity.

Each genome contributes to at most $O(m)$ ordered splits. So the sum of the cardinality of all ordered splits, i.e., the total length of all splits in Algorithm~\ref{alg}, is $O(m\,n)$. Hence, the insertion and lookups of all \texttt{S} in trie \texttt{T} takes $|\texttt{S}|$ time each and $O(m\,n)$ in total, and the number of terminal vertices of \texttt{T}, i.e., the final number of unordered splits, is in $O(m\,n)$, too.
For ease of postprocessing, splits are ordered by decreasing weight, increasing the run time for split extraction to $O(m\,n\log(m\,n))$, or $O(m\,n\log(n))$ to output only the $t$, $t\in O(n)$, highest weighting splits, respectively.

\section{Results}
\label{sec:results}
In this section, we present several use cases in order to exemplify robustness and different other characteristics of our approach SANS. We compare to the following other whole-genome based reconstruction tools.

MultiSpaM~\cite{multispam} samples a constant, high number of small, gap-free alignments of four genomes. The implied quartet topologies are combined to an overall tree topology.
To the best of our knowledge, all other tools are distance-based and rely on pairwise comparisons. Interestingly, although all methods are based on lengths or numbers of common subsequences or patterns, their results differ considerably from those of SANS.
%
%
%
Co-phylog~\cite{co-phylog} analyses each genome in terms of certain patterns \emph{(C-grams, O-grams)} and compares their characteristics \emph{(context)}.
%
In andi~\cite{andi}, enhanced suffix arrays are used to detect pairs of maximal unique matches that are used to anchor ungapped local alignments, based on which pairwise distances are computed.
CVTree3~\cite{cvtree3} corrects $k$-, $k\!-\!1$, and $k\!-\!2$-mer counts by subtracting random background of neutral mutations using a $(k\!-\!2)$-th Markov assumption.
In FSWM~\cite{fswm}, matches of patterns including match and \emph{don't-care} position are scored and filtered to estimate evolutionary distances.


Unless stated otherwise, a $k$-mer length of 31 has been used for constructing the C-DBG (Bifrost default) for SANS. Accuracy has been measured in terms of topological Robinson-Foulds distance, i.e., a predicted edge or split is correct if and only if the reference tree contains an edge that separates the same two sets of leaves. All tools have been run on a single 2\,GHz processor and times are given in CPU hours (user time).

\subsection{\textit{Drosophila}}
\label{sec:drosophila}


This dataset comprises assemblies from 12 species of the genus \textit{drosophila} obtained from the database FlyBase (\url{flybase.org}, latest release before Feb.~2019 of \url{all-chromosome}-files each)~\cite{flybase}. 

\begin{figure}
 \begin{subfigure}[t]{0.48\textwidth}
  \includegraphics[width=\textwidth]{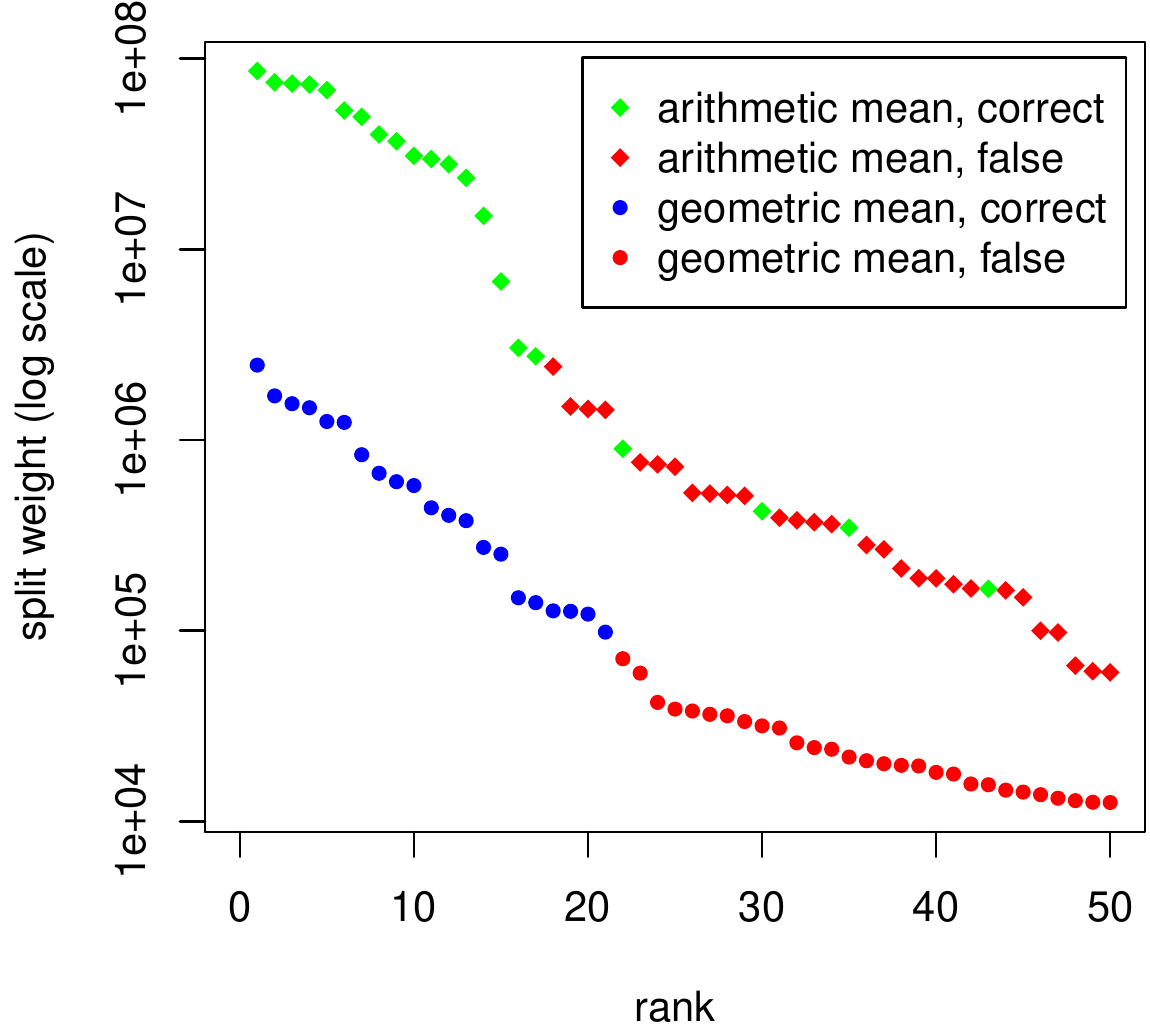}
  \caption{\label{fig:dros_add_vs_sym}Comparison of accuracy for using arithmetic or geometric mean for combining weights of splits and their inverse each. Splits have been sorted by the combined weight and the 50 highest weighting splits are shown.  Color indicates whether a split agrees with the reference~\cite{flybase}. }
 \end{subfigure}
 \hfill
 \begin{subfigure}[t]{0.48\textwidth}
  \includegraphics[width=\textwidth]{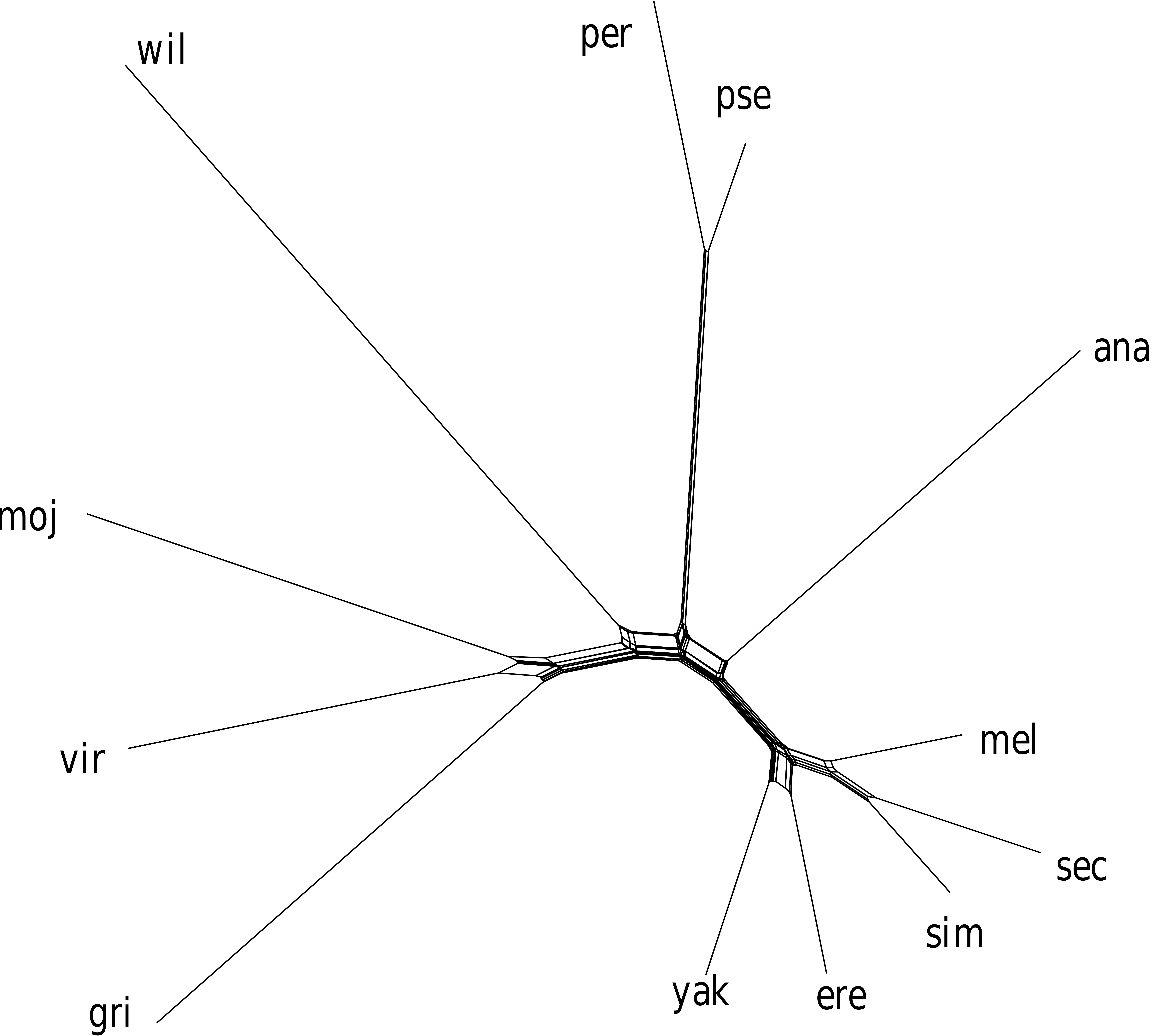}
  \caption{\label{fig:dros_net}Visualization of greedily extracted weakly compatible subset of splits using SplitsTree~\cite{splitstree_an,splitstree}. As by default, geometric mean has been used for combining weights of splits and their inverse each.}
 \end{subfigure}
 \caption{\label{fig:dros}Reconstructed phylogenetic splits on the Drosophila dataset~\cite{flybase}.}
\end{figure}

Although being ``simple'' in the sense that it contains only a small number of genomes, its analysis exemplifies the following aspects:
(i)~The effectiveness of our method for medium sized input files: for a total of more than 2\,161\,Mbp (180\,Mbp on average), SANS inferred the correct tree within 168 minutes and using up to 25\,GB of memory.
We ran CVTree3 with various values of $k$. In the best cases ($k=12$ and $13$), 7 of 9 internal edges have been inferred correctly taking 95 and 162 minutes, and up to 26 and 87\,GB of memory, respectively. (For $k=11$, only 4 internal edges were correct, and for $k>13$, the computation ran out of memory.) Both Co-phylog and FSWM did not finish within 48 hours, and both MultiSpaM and andi could not process this dataset successfully.
(ii)~As can be seen in Figure~\ref{fig:dros_add_vs_sym}, the tendency of correct splits having a high weight is stronger when combining splits and their inverse using the geometric mean than using the arithmetic mean.
(iii)~Even though the reconstruction shown in Figure~\ref{fig:dros_net} contains 45 splits---in comparison to 21 edges in a binary tree---, the visualization is close to a tree structure.

\subsection{\textit{Salmonella enterica} Para C}
\label{sec:paraC}

This dataset is of special interest as the contained assemblies from 220 genomes of different serovars within the \textit{Salmonella enterica} Para C lineage include that of an ancient Paratyphi~C genome obtained from 800 year old DNA~\cite{grapetree}, the placement of which is especially difficult due to missing data. As reference, we consider a maximum-likelihood based tree on nonrecombinant SNP data~\cite[Figure~5a]{grapetree}
.

\begin{figure}
 \begin{subfigure}[t]{0.48\textwidth}
  \includegraphics[width=\textwidth]{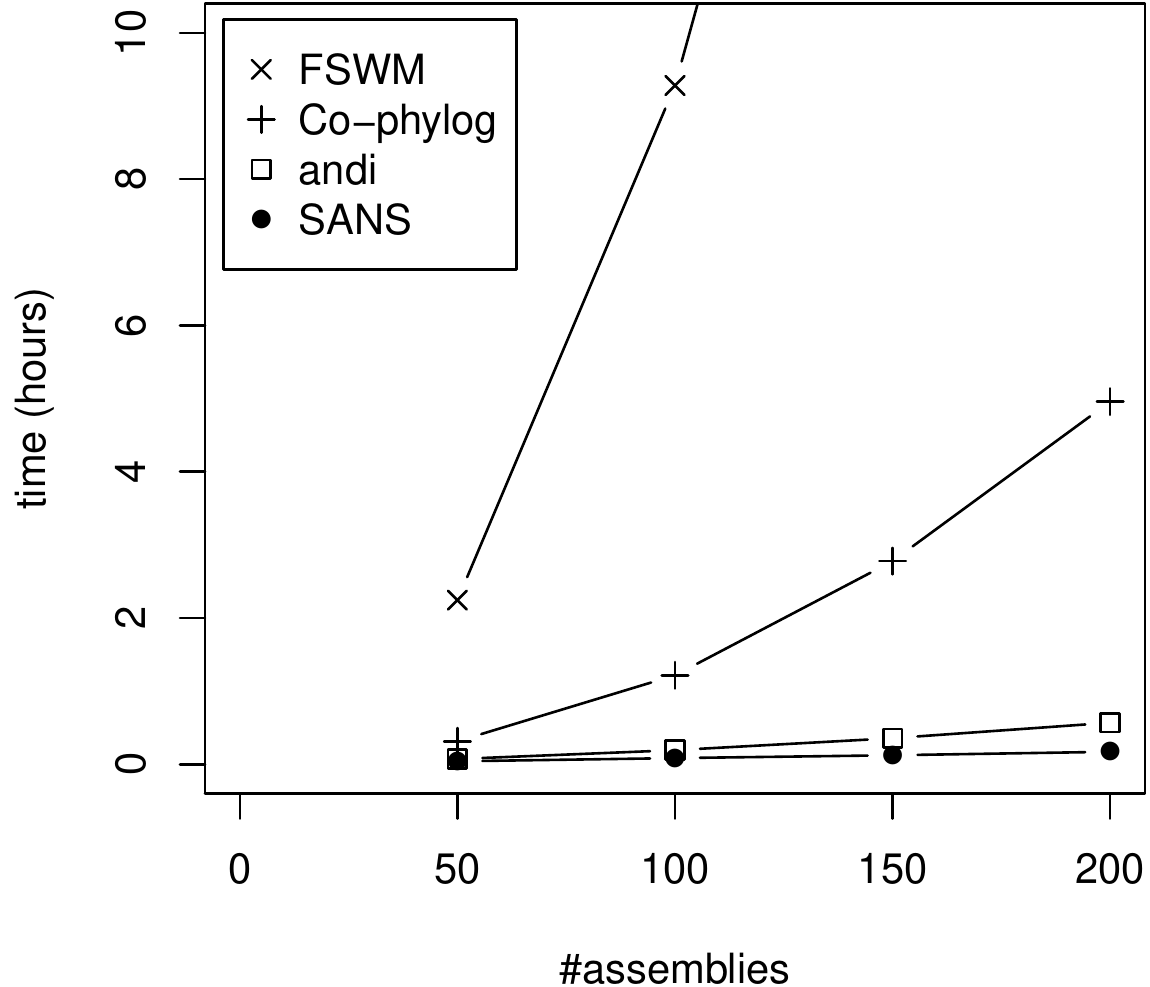}
   \caption{\label{fig:ParaC_benchmark}Running time for computing phylogenies on random subsamples. Times for SANS include DBG construction with $k=31$, split extraction and agglomeration.}
 \end{subfigure}
  \hfill
 \begin{subfigure}[t]{0.48\textwidth}
  \includegraphics[width=\textwidth]{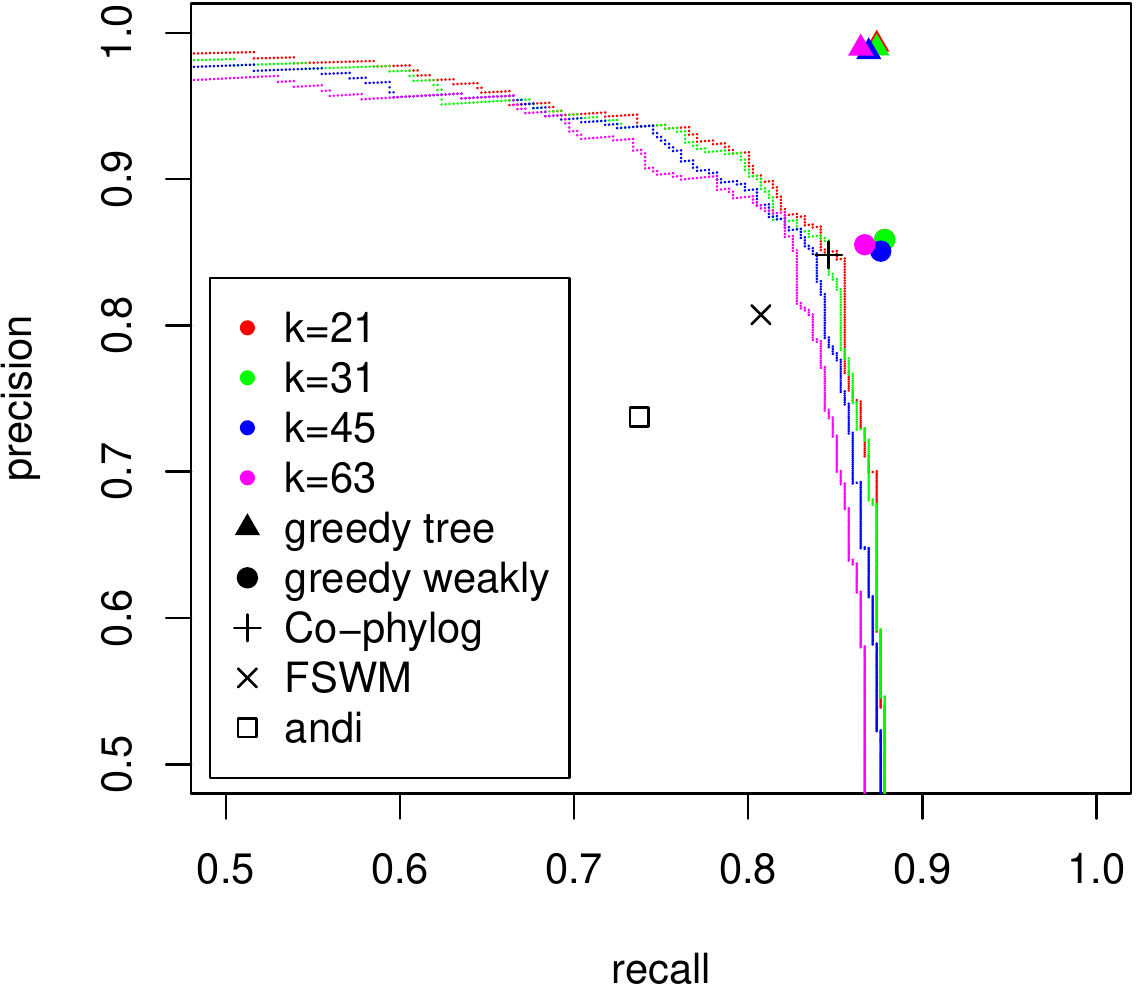}
  \caption{\label{fig:ParaC_roc_all}
  For different values of $k$,  weakly compatible subsets (triangles) and a trees (bullets) have been greedily extracted. For varying values of $i$, only the $i$ highest weighting splits have been considered as ``positives'' to determine precision and recall each (lines). 
  }
 \end{subfigure}
  \caption{\label{fig:ParaC_roc}Comparison of running time and accuracy of 
  different methods
  on the ParaC dataset~\cite{grapetree} comprising  assemblies of $n=220$ genomes.}
\end{figure}

We studied the running time behaviour of the different methods for random subsamples of increasing size. As shown in Figure~\ref{fig:ParaC_benchmark}, for this high number of closely related genomes, we observed a super-linear running time of up to 41 minutes for andi, about 5 hours for 
Co-phylog, and up to 43 hours for FSWM, whereas the reconstruction of SANS shows a linear increase (Pearson correlation coefficient 0.9994) to about 10 minutes. The memory requirement of 
both SANS and Co-phylog remained below 0.5\,GB, whereas andi required about 1\,GB, and FSWM required up to about 17\,GB.
We ran CVTree3 with ten values of $k$ between 5 and 27, but none of the resulting trees contained more than 5 correct internal edges. 
For MultiSpaM, we increased the number of sampled quartets from the default of $10^6$ to up to $10^8$, which increased the running time from about one hour to about 66 hours. Both recall and precision improved but were still below 0.2 for internal edges. 

The accuracy of the reconstructions with respect to the reference is visualized in Figure~\ref{fig:ParaC_roc_all}. In particular, we observe:
(i)~the split reconstruction by SANS and the tree inferred by Co-phylog are comparably accurate and both are more accurate than the FSWM and andi 
tree, 
(ii)~greedily extracting high weighting splits to filter for a tree selects correct splits while discarding false splits with very high precision,
(iii)~greedily extracting high weighting splits to filter for a weakly compatible subset also selects correct splits, but, as expected, has a lower precision as the tree filter, because more splits are kept than there are edges in a tree, and
(iv)~the results of SANS are robust for a wide range of $k$ from 21 to 63.


\subsection{\textit{Salmonella enterica} subspecies \textit{enterica}}

In comparison to the ParaC dataset, the 2\,964 genomes studied by Zhou \textit{et al.}~\cite{zhou2018pan} are not only a larger but also a more diverse selection of \textit{Salmonella enterica} strains. As reference, we consider a maximum-likelihood based tree on 3\,002 concatenated core genes~\cite[Figure~2A, supertree 3]{zhou2018pan}
.

The probability to observe long $k$-mers that are conserved in such a high number of more diverse genomes is lower than for the previous datasets. Hence, we selected a smaller $k$-mer length of $k=21$. To assess the efficiency and accuracy for increasing number of genomes, we sampled subsets of up to 1\,500 assemblies. To process the smallest considered subsample of size 250, andi took about 110 minutes, whereas Co-phylog and FSWM took already more than 9 and 50 hours, respectively, and MultiSpam was not able to process this dataset at all. We ran CVTree3 with all values of $k$ between 6 and 14, but in the best case ($k=8$), the resulting tree contained only 33 (of 247) correct internal edges such that we did not further consider CVTree3 in our evaluation.

\begin{figure}
 \begin{subfigure}[t]{0.52\textwidth}
  \includegraphics[width=\textwidth]{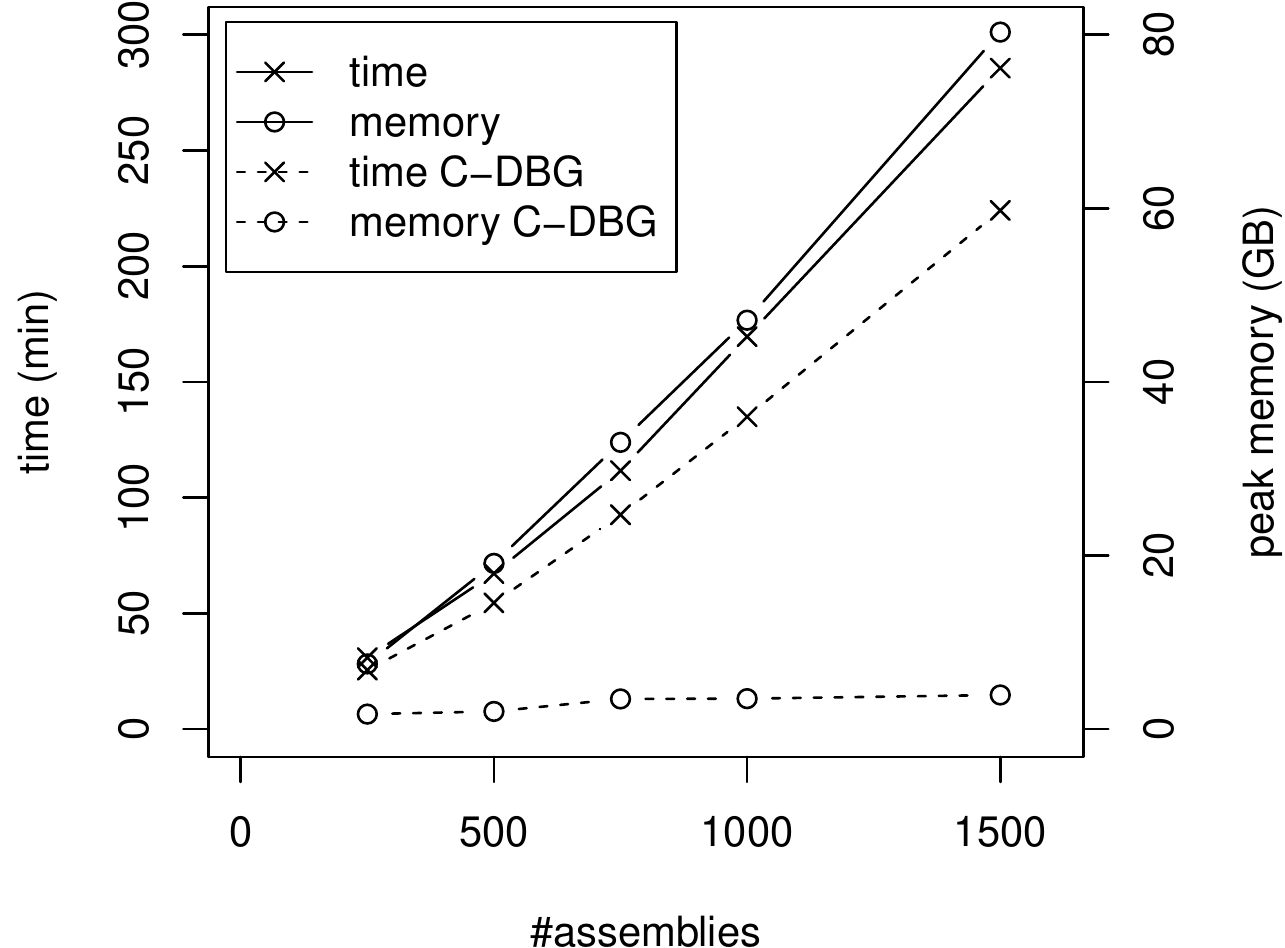}
\caption{\label{fig:3K_benchmark}Running time and peak memory usage of SANS. Values including C-DBG construction, split extraction and agglomeration, as well as C-DBG construction only are given.}
 \end{subfigure}
 \hfill
 \begin{subfigure}[t]{0.44\textwidth}
 \includegraphics[width=\textwidth]{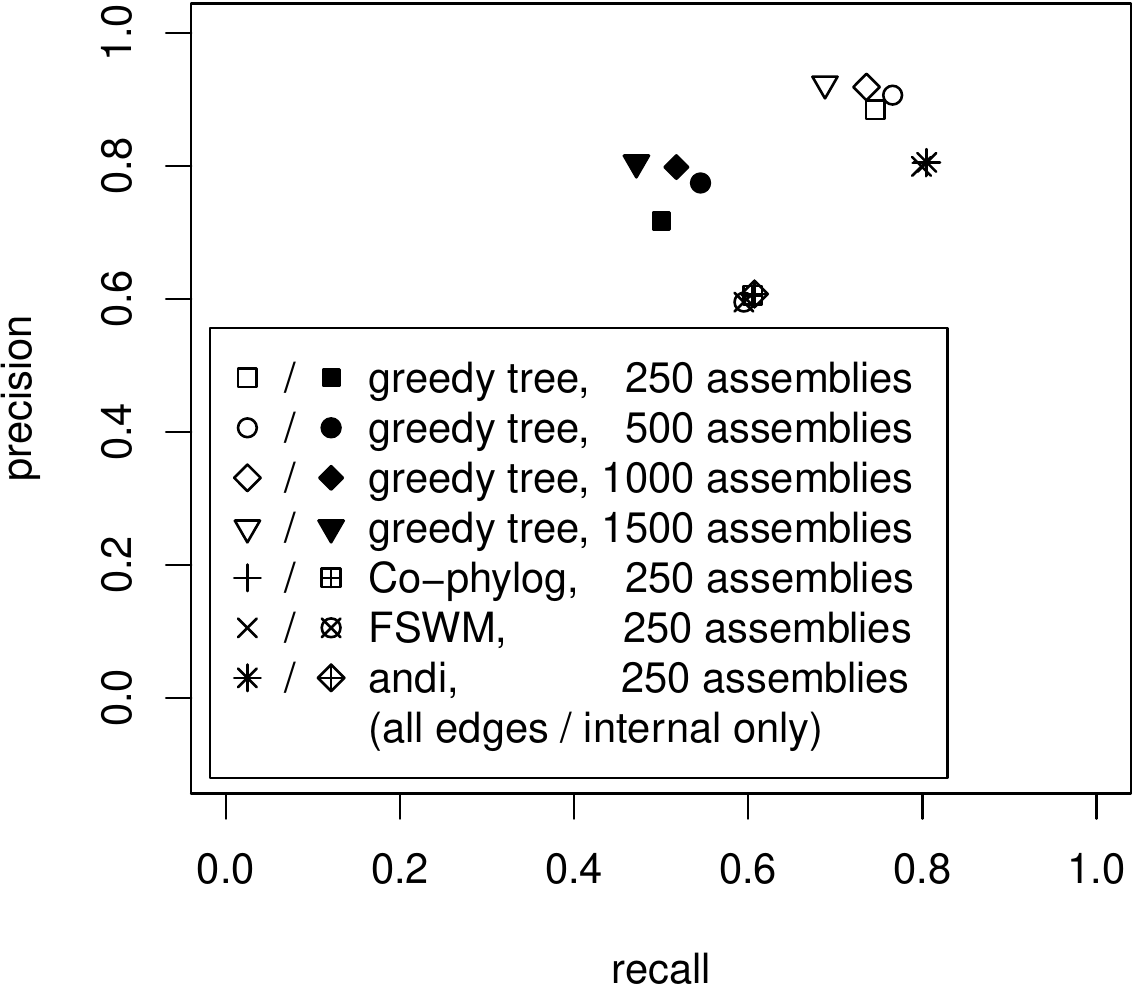}
 \caption{\label{fig:3K_roc}Accuracy with respect to the reference phylogeny~\cite[Fig.~2A]{zhou2018pan}.
  }
 \end{subfigure}
 \caption{\label{fig:3K} Efficiency and accuracy 
 on the Salmonella enterica dataset~\cite{zhou2018pan}. 
 Values have been averaged over processing two random subsamples each.
 }
\end{figure}

The memory usage for split extraction and agglomeration clearly dominates those of the C-DBG construction by Bifrost such that processing the complete dataset was not possible with our current implementation of SANS. Figure~\ref{fig:3K_benchmark} shows a slightly super-linear runtime and memory consumption of up to about 300 minutes and 80\,GB for processing 1\,500 assemblies.
As can be seen in Figure~\ref{fig:3K_roc}, both precision and recall vary only slightly for this wide range of input size. Keeping in mind that a final split of high weight strictly requires the observation of \emph{both} unordered pairs, this is a quite promising result for this first investigation of the methodology. In particular, whereas for distance-based methods, all leaf-edges are inferred by construction and can never be false, a trivial split separating a leaf from the remaining tree, requires not only some sequence unique to the leaf but also sequence that is contained in all other $n\!-\!1$ genomes. Also note that measuring accuracy by counting correct and false splits corresponding to the topological Robinson-Foulds distance has to be interpreted with care. A single misplaced leaf breaks all splits between its correct and actual location. However, this is a desired behaviour in this context, because, in a phylogeny of several hundred genomes, each genome should at least be located in the correct area, whereas the complete misplacement even of a single genome can easily lead to wrong biological conclusions.




\subsection{Prasinoviruses}

Viral genomes are short and highly diverse---posing the limits of phylogenetic reconstruction based on sequence conservation.
Here we consider complete genomes of 13 prasinoviruses, which are relatively large (213\,Kbp on average)~\cite{finke2017}. As references, we consider two trees reported in the original study, one of which is based on the presence and absence of shared putative genes~\cite[Figure~3]{finke2017}, and the other is a maximum likelihood estimation based on a marker gene (DNA polymerase B)~\cite[Figure~4]{finke2017}.

\begin{figure}
 \begin{subfigure}[t]{0.44\textwidth}
  \includegraphics[width=\textwidth]{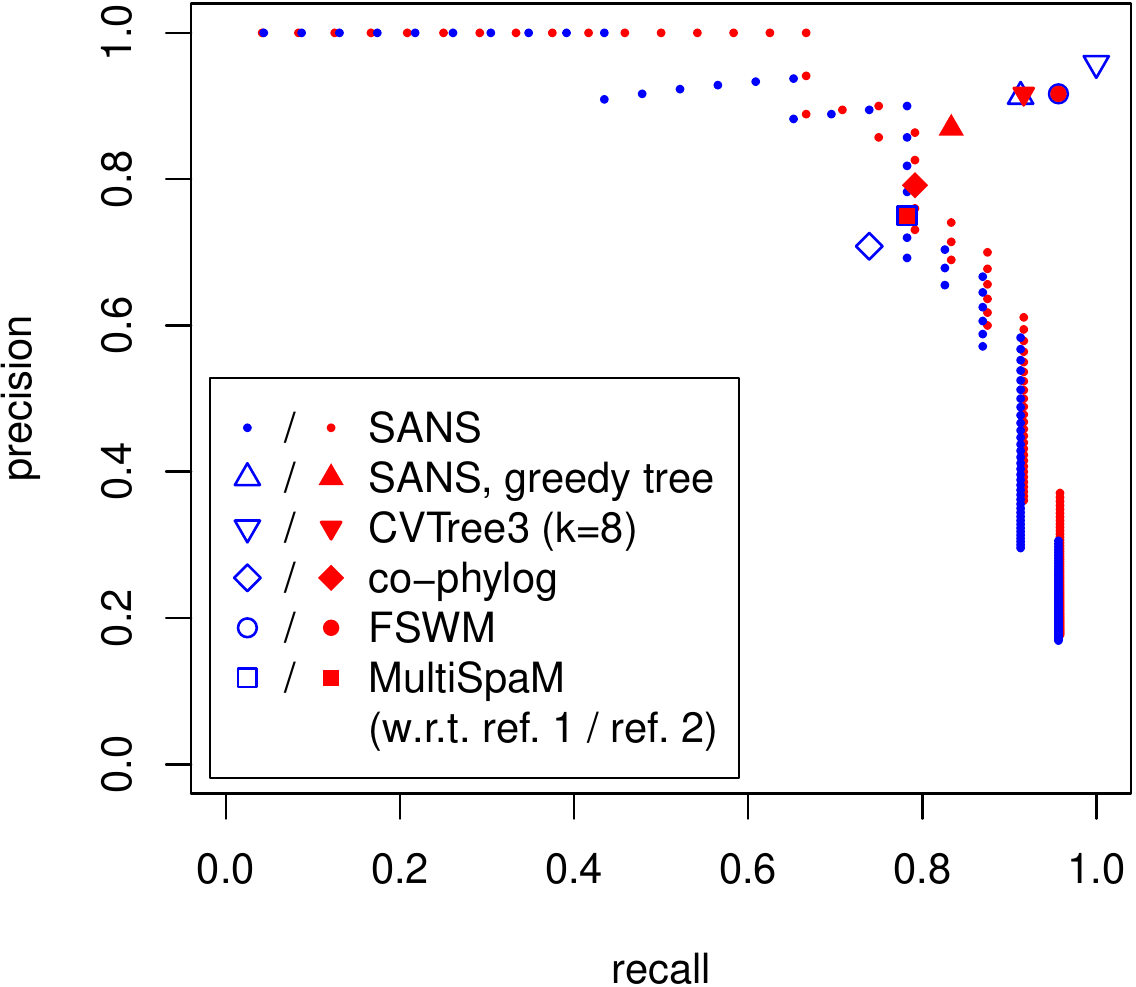}
  \caption{\label{fig:virus_roc}%
  Accuracy of different tools w.r.t.\ two reference trees~\cite[Figures 3 and 4]{finke2017} shown in blue and red, respectively. For SANS, for varying values of $i$, only the $i$ highest weighting splits have been considered as ``positives'' to determine precision and recall each
  .
  }
 \end{subfigure}\hfill%
 \begin{subfigure}[t]{0.52\textwidth}
  \includegraphics[width=\textwidth]{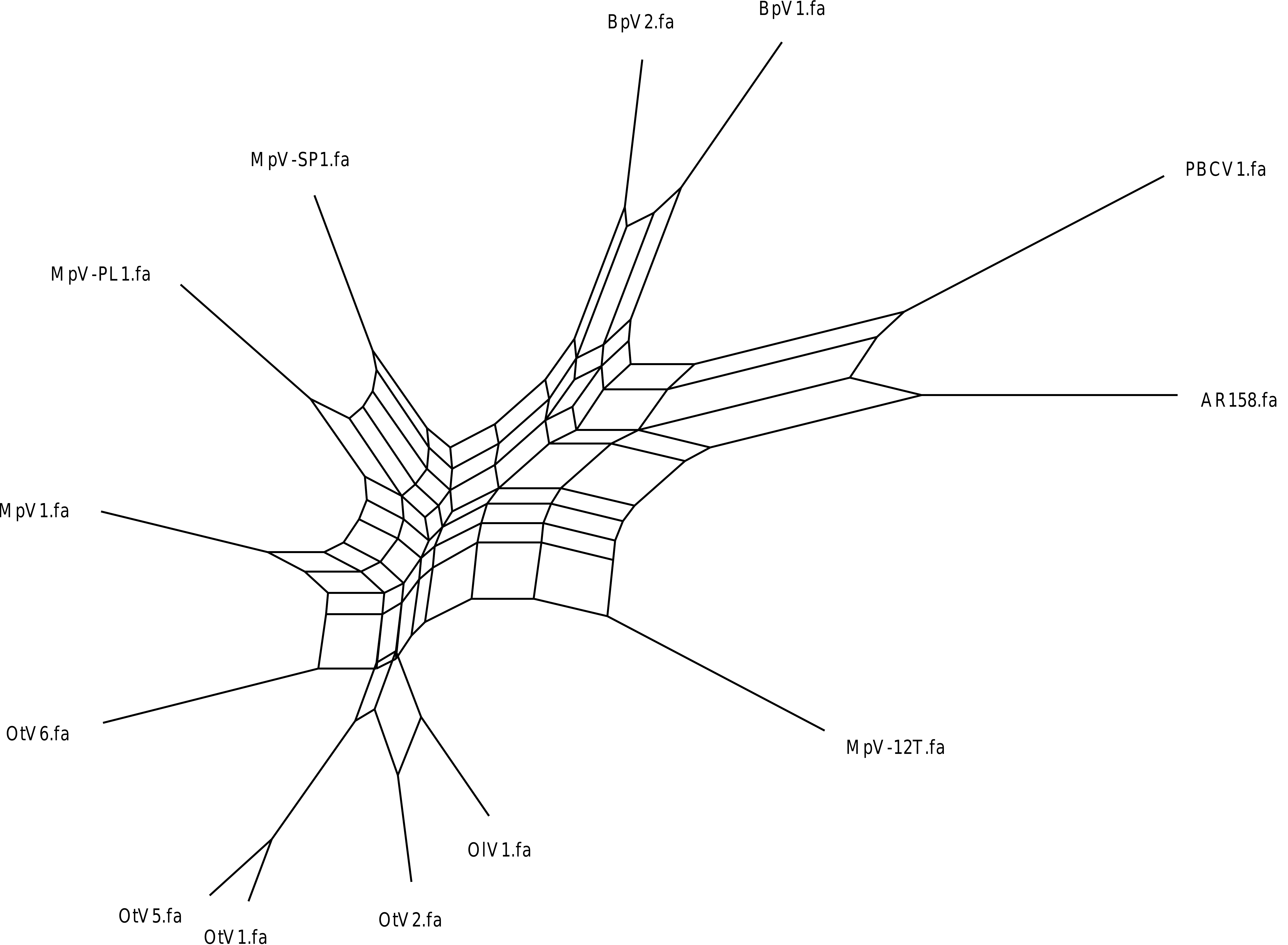}
  \caption{\label{fig:virus_net}%
  Visualization of greedily extracted weakly compatible subset of splits using SplitsTree~\cite{splitstree_an,splitstree}.
  }
 \end{subfigure}
\caption{\label{fig:virus} %
Reconstruction results on the prasinovirus dataset~\cite{finke2017}.
}\end{figure}

Due to the small size of the input, it could be processed by all tools, where time and memory consumption were negligible. Only andi could not process this dataset successfully (``very little homology was found'').
Results are shown in Figure~\ref{fig:virus_roc}.
The visualization of the predicted splits in Figure~\ref{fig:virus_net} exemplifies the explanatory power of the split framework. While main separations supported by both reference trees are recognizable as strong splits in the net, separations in which the two reference trees disagree are also shown as weakly compatible splits.

%
%
%

\subsection{\textit{Vibrio cholerae}}

The dataset comprises 22 genomes from the species \textit{Vibrio  cholerae}, 7 of which have been sequenced from clinical samples and are labelled ``pandemic  genome'' (PG), and the remaining 15 have been sequenced from non-clinical samples and are labelled ``environmental genome'' (EG)~\cite[primary dataset]{shapiro2017}.
As already observed in the original study, for these genomes, it is difficult to reconstruct a reliable, fully resolved tree. Nevertheless, representing the phylogeny in form of splits shows a strong separation of the pandemic from the environmental group.
The phylogeny presented by the authors of the original study~\cite[Supplementary Figure~1a]{shapiro2017} is based on 126\,099 
sites extracted from alignment blocks.

\begin{figure}
 \begin{subfigure}[t]{0.58\textwidth}
  \includegraphics[width=\textwidth]{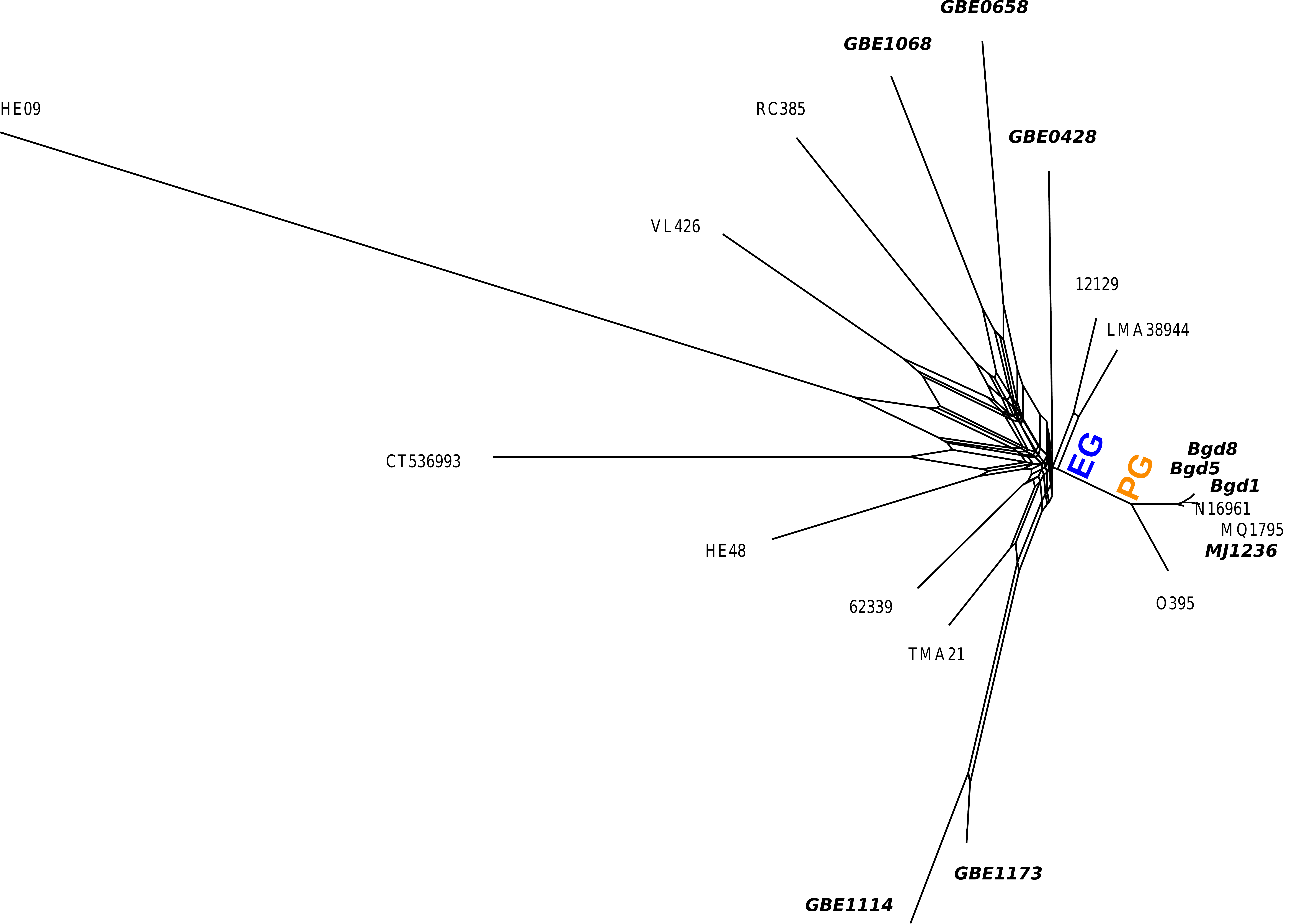}
  \caption{\label{fig:cholerae_splits} Visualization of greedily extracted weakly compatible subset of splits. For taxa highlighted in bold, only read data was available on NCBI (input option \texttt{-s} of Bifrost has been used); for Taxon TM1107980, no data was available on NCBI (February 2019).}
 \end{subfigure}
 \hfill
 \begin{subfigure}[t]{0.38\textwidth}
  \includegraphics[width=\textwidth]{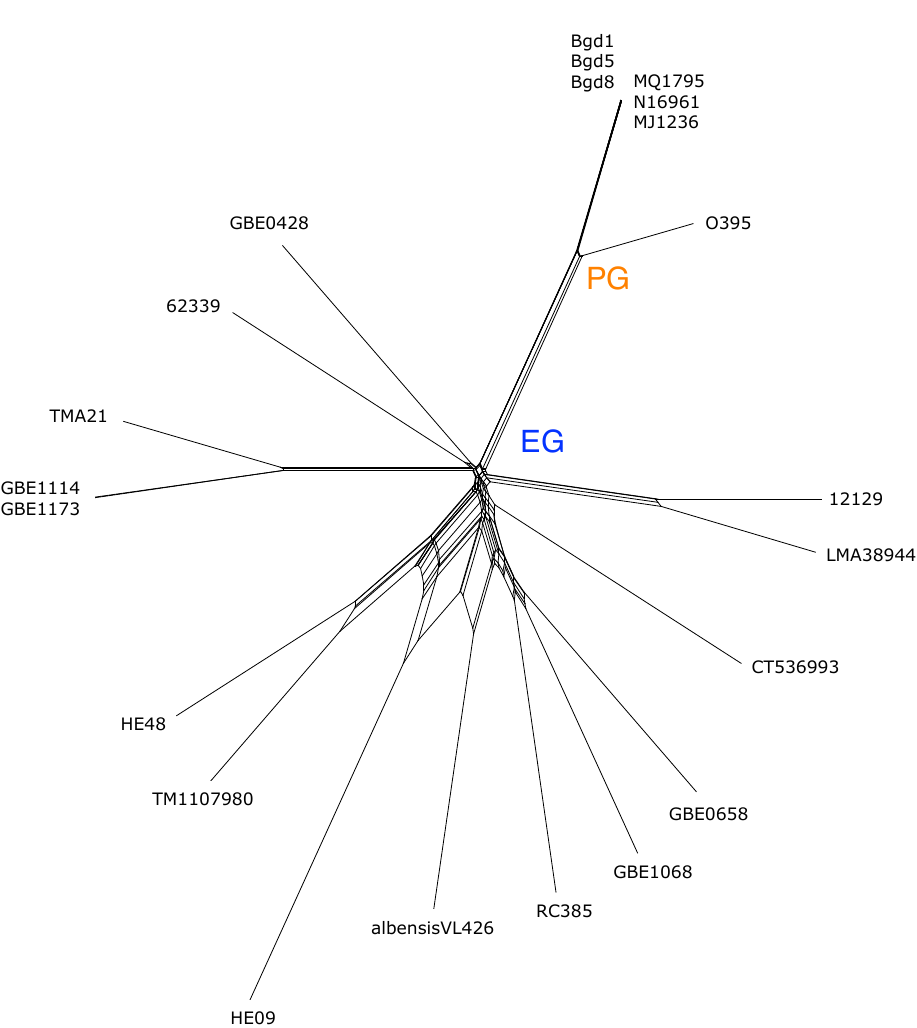}
  \caption{\label{fig:cholerae_ref}
  Reference phylogeny. Figure reprinted from Shapiro \textit{et al.}~\cite[Supplementary Figure~1a]{shapiro2017}.}
 \end{subfigure}
 \caption{\label{fig:cholerae}Splits reconstructed for the \textit{V.~cholerae} dataset~\cite{shapiro2017} by SANS (left) and by Shapiro \textit{et al.}~\cite{shapiro2017} (right) visualized with SplitsTree~\cite{splitstree_an,splitstree}.}
\end{figure}

Comparing our reconstruction results to the reference, both shown in Figure~\ref{fig:cholerae}, we make two observations.
(i)~Our reconstruction also separates the pandemic from the environmental group, and agrees to the reference in further sub-groups.
(ii)~When collecting the sequence data, for some of the genomes, we found assemblies, whereas for others, only read data was available. Because the used C-DBG implementation Bifrost supports a combination of both types as input, we were able to reconstruct a joint phylogeny without extra effort or obvious bias in the result.

\section{Discussion and Outlook}
\label{sec:conclusion}

We proposed a new $k$-mer based method for phylogenetic inference that neither relies on alignments to a reference sequence nor on pairwise or multiple alignments to infer markers. 
Prevailing whole-genome approaches perform pairwise comparisons to determine a quadratic number of distances to finally infer a linear number of tree edges. In contrast, in our approach, the length of conserved sequences is extracted from a colored de-Bruijn graph to first infer signals for phylogenetic sub-groups. These signals are then combined with a symmetry assumption to weighted phylogenetic splits.
Evaluations on several real datasets have proven comparable or better efficiency and accuracy compared to other whole-genome approaches. Our results indicate robustness in terms of $k$-mer length, as well as the taxonomic order, size and number of the genomes. The analysis of a dataset composed of both assembly and read data indicated also robustness in this regard---an important feature, which we want to investigate further.

A distinctive feature of the proposed methodology is the direct association of a phylogenetic split to the conserved subsequences it has been derived from, which is not possible for distance-based methods. We plan to enrich our implementation with this valuable possibility to allow the analysis of characteristic subsequences of identified subgroups, or subsequences inducing phylogenetic splits off the main tree, e.g.~horizontal gene transfer.
Here, the applied generalization of trees plays an important role, e.g., circular split systems are more strict than weakly compatible sets and might thus be a promising alternative to be studied further.

Finally, we want to emphasize the simplicity of the new approach as presented here. At its current state, apart from iterating a colored de-Bruijn graph and agglomerating unitig  lengths, the only elaborate ingredient so far is the symmetry assumption realized by applying the geometric mean. We believe that the general approach still harbors much potential to be further refined by, e.g., statistical models, advanced data structures, pre- or postprocessing, to further increase its accuracy and efficiency.

\section*{Acknowledgements}
I thank Guillaume Holley for support on Bifrost, Nina Luhmann for pointers to data sets, and Andreas Rempel for programming assistance.

\bibliography{references}


\end{document}